\documentclass[fleqn,twoside]{article}
\usepackage{espcrc2,cite,graphicx,amsmath,amsfonts,amssymb,mathrsfs,epic,eepic}

\newcommand{\ie}{{\it i.e.}}

\newcommand{\eg}{{\it e.g.}}

\newcommand{\cf}{{\it cf.}}

\newcommand{\eq}{Eq.}

\newcommand{\fig}{Fig.}
\newcommand{\figs}{Figs.}
\newcommand{\Ref}{Ref.}
\newcommand{\Refs}{Refs.}
\newcommand{\Sec}{Sec.}

\begin{document}

\title{
\vspace*{-3cm}
\begin{flushright}
{\small TUM-HEP-433/01}
\end{flushright}
{\bf Effects and influences on neutrino oscillations due to a thin
density layer perturbation added to a matter density profile}}

\author{{\large Tommy Ohlsson}\address[TUM]{{\it Institut f{\"u}r Theoretische
Physik, Physik-Department, Technische Universit{\"a}t M{\"u}nchen,
James-Franck-Stra{\ss}e, 85748 Garching bei M{\"u}nchen,
Germany}}\thanks{E-mail address: {\tt tohlsson@ph.tum.de}}}
     
\begin{abstract}
\noindent {\bf Abstract} \vspace{2.5mm}

In this paper, we show the effects on the transition probabilities for
neutrino oscillations due to a thin constant density layer perturbation
added to an arbitrary matter density profile. In the case of two
neutrino flavors, we calculate the effects both analytically and
numerically, whereas in the case of three neutrino flavors, we perform
the studies purely numerically. As an realistic example we consider
the effects of the Earth's atmosphere when added to the Earth's matter
density profile on the neutrino oscillation transition probabilities
for atmospheric neutrinos.

\vspace*{0.2cm}
\noindent {\it PACS:} 14.60.Lm; 13.15.+g\\
\noindent {\it Key words:} Neutrino oscillations; Matter effects; Atmospheric
neutrinos; Earth's atmosphere and matter density profile
\end{abstract}

\maketitle

\section{Introduction}

The measurements in neutrino oscillation experiments are becoming more
and more accurate (\eg{} the Super-Kamiokande and SNO experiments
\cite{SuperK,SNO}). This
means that we have to use three neutrino flavors in the analyses of
data. The measurement of atmospheric neutrinos contains also another
important issue. How large are the effects on the neutrino oscillation
probabilities due to the Earth's atmosphere?
In order to answer this question, we will in this paper in general consider the
effects of a thin density layer perturbation on an arbitrary matter
density profile.

This paper is organized as follows: In \Sec~\ref{sec:form}, we derive
the formalism. In \Sec~\ref{sec:23nu}, the special
cases of two and three neutrino flavors are investigated as well as
the analysis of atmospheric neutrinos. Finally, in
\Sec~\ref{sec:sc}, we present a summary as well as our conclusions.

\section{Formalism}
\label{sec:form}

Neutrino propagation in matter of constant density can be described by
an evolution operator
\begin{equation}
U_f(L;A) = e^{-i \mathscr{H}_f(A) L},
\end{equation}
where $\mathscr{H}_f(A) \equiv U H_m U^{-1} + A \; {\rm
diag\,}(1,0,\ldots,0)$ is the total Hamiltonian in the flavor basis,
$L$ is the neutrino (traveling) path length, \ie, the baseline length,
and $A$ is the (constant) matter density parameter related to the
(constant) matter density $\rho$
as $A = \sqrt{2} G_F \frac{Y_e}{m_N} \rho$. Here $H_m \equiv
{\rm diag\,}(E_1,E_2,\ldots,E_n)$, where $E_a = m_a^2/(2E_\nu)$, is
the free Hamiltonian in the mass basis, $U = (U_{\alpha a})$ is the
leptonic mixing matrix, $G_F \simeq 1.16639 \cdot 10^{-23} \; {\rm
eV}^{-2}$ is the Fermi weak coupling constant, $Y_e$ is the average
number of electrons per nucleon\footnote[1]{Earth: $Y_e \simeq
\frac{1}{2}$.}, and $m_N \simeq 939 \;
{\rm MeV}$ is the nucleon mass. Furthermore, $n$ is the number of
neutrino flavors, $m_a$ is the mass of the $a$th mass eigenstate, and
$E_\nu$ is the neutrino energy.

Using the evolution operator method developed in \Ref \cite{ohls002},
the total evolution operator in matter consisting of $N$ different
(constant) matter density layers is given by 
\begin{equation}
U_f(L) = \prod_{k=1}^N U_f(L_k;A_k),
\end{equation}
where $L_k$ and $A_k$ is the thickness and matter density parameter of
the $k$th matter density layer, respectively, and $L \equiv \sum_{k=1}^N L_k$.
Similar methods to the evolution operator method for propagation of
neutrinos in matter consisting of two density layers using two
neutrino flavors have been discussed in \Refs~\cite{petc98,akhm99}.

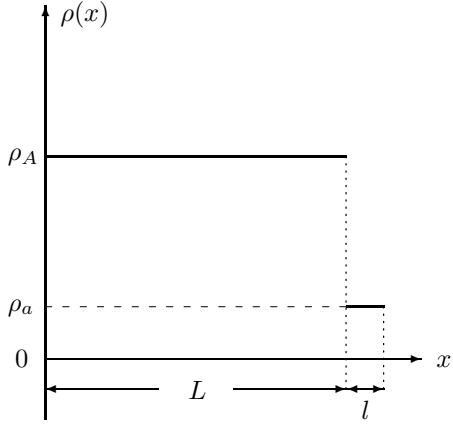
\begin{figure}
\begin{picture}(6,5)
\put(1,-0.5){\vector(0,1){5.5}} \put(1.2,4.8){$\rho(x)$}
\put(1,0.3){\vector(1,0){5}} \put(6.2,0.2){$x$} \put(0.6,0.2){0}
\thicklines
\put(1,3){\line(1,0){4}}
\put(5,1){\line(1,0){0.5}}
\thinlines
\put(2.5,-0.1){\vector(-1,0){1.5}} \put(3.5,-0.1){\vector(1,0){1.5}}
\put(2.9,-0.225){$L$}
\put(5.25,-0.1){\vector(-1,0){0.25}} \put(5.25,-0.1){\vector(1,0){0.25}}
\put(5.2,-0.5){$l$}
\put(0.5,2.95){$\rho_A$} \put(0.5,0.95){$\rho_a$}
\dottedline{0.1}(5,-0.1)(5,3) \dottedline{0.1}(5.5,-0.1)(5.5,1)
\dashline{0.1}(1,1)(5,1)
\end{picture}
\caption{The (constant) matter density profile with the thin
(constant) density layer perturbation. The densities $\rho_A$ and $\rho_a$
correspond to the matter density parameters $A$ and $a$,
respectively. Note that $\rho_a$ can in general be both smaller and
larger than $\rho_A$.}
\label{fig:density}
\end{figure}

In order to investigate the effects of a thin density layer perturbation on
a constant matter density profile, we can thus use this method and we
obtain the total evolution operator as
\begin{equation}
U_f(L') = U_f(l;a) U_f(L;A),
\label{eq:evol_pert}
\end{equation}
where $L' = L + l$. Here $l$ and $a$ are the path length and matter density
parameter of the density perturbation, respectively. See
\fig~\ref{fig:density} for the geometry. Without loss of
generality, we can in fact set $a = 0$ and we will write
Eq.~(\ref{eq:evol_pert}) as \cite{chiz01}
\begin{equation}
U_f(L') = U_f(l) U_f(L;A).
\label{eq:evol_pert_2}
\end{equation}
Note that it is the density difference $\Delta \rho \equiv \rho_A -
\rho_a$ that is important, \ie, the absolute density scale is not
crucial, which means that one can choose $a = 0$.

The neutrino oscillation transition amplitude from a neutrino flavor
$\nu_\alpha$ to a neutrino flavor $\nu_\beta$ is defined as
\begin{equation}
A_{\alpha\beta} \equiv \langle \nu_\beta \vert U_f \vert
\nu_\alpha \rangle, \quad \alpha,\beta = e,\mu,\tau, \ldots.
\label{eq:A}
\end{equation}
Therefore, the corresponding neutrino oscillation transition probability
for $\nu_\alpha \leftrightarrow \nu_\beta$ is given by
\begin{equation}
P_{\alpha\beta} \equiv \left\vert A_{\alpha\beta} \right\vert^2, \quad
\alpha,\beta = e,\mu,\tau,\ldots.
\label{eq:P}
\end{equation}
Inserting Eq.~(\ref{eq:evol_pert_2}) into Eq.~(\ref{eq:A}) and the
result thereof into Eq.~(\ref{eq:P}), we find
\begin{equation}
P_{\alpha\beta}(L,l;A) = \left\vert \langle \nu_\beta \vert U_f(l)
U_f(L;A) \vert \nu_\alpha \rangle \right\vert^2,
\end{equation}
which can be re-written as
\begin{equation}
P_{\alpha\beta}(L,l;A) = \left\vert \sum_\gamma \langle \nu_\beta
\vert U_f(l) \vert \nu_\gamma \rangle \langle \nu_\gamma \vert
U_f(L;A) \vert \nu_\alpha \rangle \right\vert^2.
\end{equation}
Expanding this equation gives
\begin{eqnarray}
&& P_{\alpha\beta}(L,l;A) = \sum_\gamma P_{\alpha\gamma}(L;A)
P_{\gamma\beta}(l) \nonumber\\
&+& \underset{\gamma \neq \delta}{\sum_\gamma \sum_\delta}
A^\ast_{\alpha\gamma}(L;A) A^\ast_{\gamma\beta}(l)
A_{\alpha\delta}(L;A) A_{\delta\beta}(l). \nonumber\\\label{eq:PLlA}
\end{eqnarray}
Note that the summation indices in the sums run over all $n$ flavor
states.

As measures of the effects of the thin density layer perturbation, we will
define the following difference
\begin{equation}
\Delta P_{\alpha\beta} \equiv P_{\alpha\beta}(L,l;A) -
P_{\alpha\beta}(L;A),
\label{eq:DP}
\end{equation}
which corresponds to the absolute error of the neutrino oscillation
transition probability $P_{\alpha\beta}$, as well as the two relative
errors
\begin{equation}
\epsilon_{\alpha\beta} \equiv \frac{\Delta
P_{\alpha\beta}}{P_{\alpha\beta}(L,l;A)}
\end{equation}
and
\begin{equation}
\epsilon'_{\alpha\beta} \equiv \frac{\Delta
P_{\alpha\beta}}{P_{\alpha\beta}(L;A)}.
\end{equation}
Note that it holds that $\epsilon_{\alpha\beta} \approx
\epsilon'_{\alpha\beta}$.

\section{Two and three neutrino flavors}
\label{sec:23nu}

In this section, we will discuss the particular cases of two and three
neutrino flavors.

For two neutrino flavors we have for the thin constant density layer
perturbation \cite{ohls00}
\begin{eqnarray}
P_{\alpha\beta}(l) &=& \delta_{\alpha\beta} \nonumber\\
&-& 4 \; \underset{a<b}{\sum_{a=1}^2\sum_{b=1}^2} U_{\alpha a} U_{\beta a}
U_{\alpha b} U_{\beta b} \sin^2 x_{ab}, \nonumber\\\label{eq:Pl}\\
A_{\alpha\beta}(l) &=& \sum_{a=1}^2 U_{\alpha a} U_{\beta a} e^{-i E_a
l}, \label{eq:Al}
\end{eqnarray}
where $x_{ab} \equiv \frac{(E_a - E_b)l}{2}$ and the leptonic mixing
matrix elements are given in terms of the vacuum mixing angle $\theta$
as $U_{e1} = U_{\mu 2} = \cos \theta$ and $U_{e2} = - U_{\mu 1} = \sin \theta$.
Inserting Eqs.~(\ref{eq:Pl}) and (\ref{eq:Al}) into
Eq.~(\ref{eq:PLlA}) and the result thereof into Eq.~(\ref{eq:DP}), we
obtain (for $\alpha = e$ and $\beta = \mu$) after some tedious calculations
\begin{eqnarray}
&& \Delta P_{e\mu}(L,l;A) = \sin^2 2\theta \left[ 1 - 2 P_{e\mu}(L;A)
\right] \nonumber\\
&\times& \sin^2 x_{21} + \sin 2\theta \; \Re \big\{ A^\ast_{ee}(L;A)
A_{e\mu}(L;A) \nonumber\\
&\times& \left[ \cos 2\theta - \left( e^{-2i x_{21}} \cos^2 \theta -
e^{2i x_{21}} \sin^2 \theta \right) \right] \big\}, \nonumber\\
\end{eqnarray}
where $\Re z$ denotes the real part of the complex number $z$.
Furthermore, we have for the constant matter density profile (matter
density parameter $A \neq 0$) \cite{ohls00}
\begin{eqnarray}
P_{e\mu}(L;A) &=& \frac{\sin^2 2\theta}{\sin^2 2\theta + \left( \cos
2\theta - \frac{AL}{2 X_{21}} \right)^2} \sin^2 \Omega L,
\nonumber\\\\
A_{ee}(L;A) &=& e^{i\phi} \bigg[ \cos \Omega L \nonumber\\
&+& i \frac{X_{21}}{\Omega L} \left( \cos
2\theta - \frac{AL}{2 X_{21}} \right) \sin \Omega L \bigg],\nonumber\\\\
A_{e\mu}(L;A) &=& -i e^{i\phi} \frac{X_{21}}{\Omega L} \sin 2\theta \sin \Omega L,
\end{eqnarray}
where $X_{21} \equiv \frac{\Delta m^2 L}{4E_\nu}$ (\cf{} $x_{21} =
\frac{\Delta m^2 l}{4E_\nu}$),
$$
\Omega \equiv \sqrt{\frac{\Delta m^2}{4E_\nu} \left(
\frac{\Delta m^2}{4E_\nu} - A \cos 2\theta \right) + \frac{A^2}{4}},
$$
and $e^{i\phi}$ is a phase factor. ($\left. \Omega \right\vert_{A = 0} =
\frac{\Delta m^2}{4E_\nu} = \frac{X_{21}}{L}$) Using these relations,
we finally obtain
\begin{eqnarray}
\Delta P_{e\mu}(L,l;A) &=& \sin^2 2\theta \bigg\{ \sin^2 x_{21} \bigg[
1 + 2 \left( \frac{X_{21}}{\Omega L} \right)^2 \nonumber\\
&\times& \left( \frac{AL}{2 X_{21}} \cos 2\theta - 1 \right)
\sin^2 \Omega L \bigg] \nonumber\\
&+& \frac{1}{2} \frac{X_{21}}{\Omega L} \sin 2 x_{21} \sin 2 \Omega L
\bigg\} \nonumber\\\label{eq:DP2}
\end{eqnarray}
and
\begin{eqnarray}
\epsilon'_{e\mu}(L,l;A) &=& \left(\frac{\Omega L}{X_{21}}\right)^2
\frac{1}{\sin^2 \Omega L} \nonumber\\
&\times& \bigg\{ \sin^2 x_{21} \bigg[
1 + 2 \left( \frac{X_{21}}{\Omega L} \right)^2 \nonumber\\
&\times& \left( \frac{AL}{2 X_{21}} \cos 2\theta - 1 \right)
\sin^2 \Omega L \bigg] \nonumber\\
&+& \frac{1}{2} \frac{X_{21}}{\Omega L} \sin 2 x_{21} \sin 2 \Omega L
\bigg\}. \nonumber\\\label{eq:e2}
\end{eqnarray}
So far we have carried out all calculations exactly.
Note that for $A = 0$ we have $P_{e\mu}(L,l;0) = \sin^2 2\theta (
\sin^2 x_{21} + \sin^2 X_{21} + \frac{1}{2} \sin 2x_{21} \sin 2X_{21}
- 2 \sin^2 x_{21} \sin^2 X_{21} ) = \sin^2 2\theta \sin^2 (x_{21} +
X_{21}) = P_{e\mu}(L+l;0)$, since $\Delta P_{e\mu}(L,l;0) =
P_{e\mu}(L,l;0) - P_{e\mu}(L;0)$ and $P_{e\mu}(L;0) = \sin^2 2\theta
\sin^2 X_{21}$.

Let us now assume that $l \ll L$, which also implies that $x_{21} \ll
X_{21}$. This means that we can series expand Eqs.~(\ref{eq:DP2}) and
(\ref{eq:e2}) in the small dimensionless parameter $x_{21} = X_{21}
\frac{l}{L}$. Carrying out the series expansions, we obtain to first
order in $\frac{l}{L}$
\begin{eqnarray}
\Delta P_{e\mu}(L,l;A) &\simeq& \sin^2 2\theta \sin 2 \Omega L
\frac{X_{21} x_{21}}{\Omega L} \nonumber\\
&=& \sin^2 2\theta \sin 2 \Omega L
\frac{\Delta m^4 L}{4 \Omega E_\nu^2} \frac{l}{L} \nonumber\\
\end{eqnarray}
and
\begin{equation}
\epsilon'_{e\mu}(L,l;A) \simeq 2 \Omega L \cot \Omega L \frac{l}{L}.
\label{eq:e'em}
\end{equation}

As a realistic example, let us now discuss the influence of the
Earth's atmosphere on the transition probabilities for atmospheric
neutrinos traversing the Earth. The atmosphere will in this case serve
as the thin constant density layer perturbation. To a very good
approximation the density of the atmosphere can be assumed to be equal
to zero ($\rho_a = 0$). Note that atmospheric neutrinos first traverse
the atmosphere and then the Earth's matter density profile, \ie, they
first traverse the thin constant density layer perturbation and then
the matter density profile, which is the reverse total matter density
profile to the one shown in \fig~\ref{fig:density}. In the case of two neutrino
flavors, the direct and reverse total matter density profiles will
give the same results, since there are no T-violating effects in this
case ($P_{e\mu} = P_{\mu e} \quad \Rightarrow \quad \Delta P_{e\mu}^T
\equiv P_{e\mu} - P_{\mu e} = 0$). However, in the case
of three neutrino flavors, the order of the thin constant density
layer perturbation and the matter density profile plays a role and
will cause a non-zero matter-induced T violation \cite{fish01,akhm012}. The
traveling path lengths of atmospheric neutrinos in the Earth's matter
density profile 
and the Earth's atmosphere 
can be found from simple geometrical considerations and are given by
\begin{eqnarray}
L \equiv L_\oplus(h) &=& 2 R_\oplus \cos h, \\
l \equiv L_{\rm atm.}(h) &=& - R_\oplus \cos h \nonumber\\
&+& \sqrt{(R_\oplus + d)^2 - R_\oplus^2 \sin^2 h}, \nonumber\\
\end{eqnarray}
respectively, where $R_\oplus \simeq 6371 \; {\rm km}$ is the
equatorial radius
of the Earth, $d = (10 - 20) \; {\rm km} \sim 15 \; {\rm km}$ is the
thickness of the Earth's atmosphere, and $0 \leq h \leq 90^\circ$ is
the nadir angle.

\begin{figure}
\includegraphics*[bb = 0 0 590 700,height=7.5cm,angle=270]{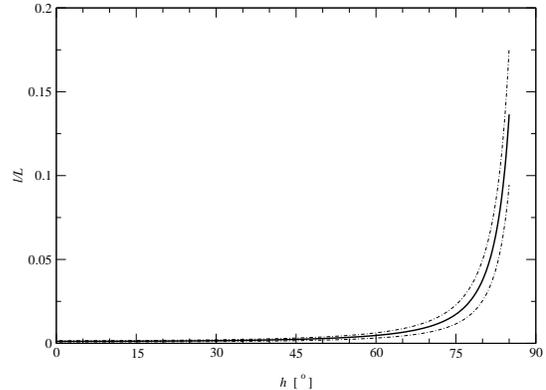}
\vspace{-1cm}
\caption{The ratio between the distances traveled by neutrinos in the
Earth's atmosphere and the Earth's matter density profile $l/L$ as a
function of the nadir angle $h$. The solid curve is plotted for $d = 15 \;
{\rm km}$, whereas the two dashed-dotted curves are plotted for $d =
10 \; {\rm km}$ and $d = 20 \; {\rm km}$, respectively.} 
\label{fig:lL}
\end{figure}

The ratio of traveling path lengths in the atmosphere and in the
Earth,
$$
\frac{l}{L} = \frac{1}{2} \left( \sqrt{\left(1 + \tfrac{d}{R_\oplus}\right)^2
\left(1 + \tan^2 h\right) - \tan^2 h} - 1 \right),
$$
is plotted as a function of the nadir angle $h$ in
\fig~\ref{fig:lL}.
Since we are interested in the traveling paths of
atmospheric neutrinos that are only a small fraction in the
atmosphere, we will restrict ourselves to the cases when $l/L \lesssim
0.10$, which corresponds to a nadir angle
$$
h = \arctan \sqrt{\frac{(1 + 2 \frac{l}{L} )^2 - (1 +
\frac{d}{R_\oplus})^2}{(1 + \frac{d}{R_\oplus})^2 - 1}} \lesssim 84^\circ.
$$
In the limit $h \to 90^\circ$, the neutrinos travel the ``maximal''
distance in the atmosphere and just touch the surface of the Earth.

For two neutrino flavors we have the effective (vacuum) mixing
parameters $\theta \simeq \theta_{13} \simeq 5^\circ$, which is below
the CHOOZ upper bound\footnote{The CHOOZ upper bound is: $\sin^2 2
\theta_{13} \lesssim 0.10$ \cite{CHOOZ,CHOOZ2} \quad $\Rightarrow$
\quad $\theta_{13} \lesssim 9.2^\circ$.} and $\Delta m^2 \simeq
\Delta m_{32}^2 \simeq 2.5 \cdot 10^{-3} \; {\rm eV}^2$
\cite{SuperK}, which is the latest best fit value of the
Super-Kamiokande collaboration. Furthermore, we assume completely
Earth-through-going atmospheric neutrinos ($h = 0$) with neutrino energy
$E_\nu = 1 \; {\rm GeV}$, which is a reasonable value of the neutrino
energy for atmospheric neutrinos. 

\begin{figure}
\includegraphics*[bb = 0 0 590 700,height=7.5cm,angle=270]{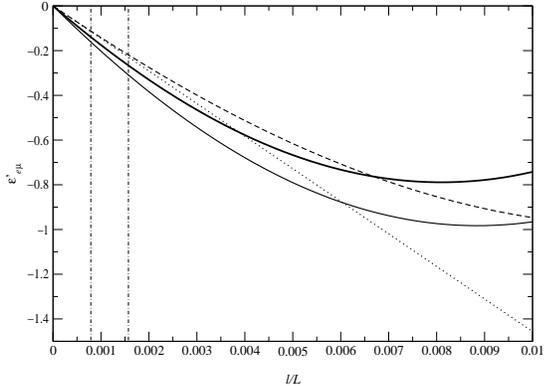}
\vspace{-1cm}
\caption{The relative error $\epsilon'_{e\mu}$ as a function of the
ratio $l/L$ for atmospheric neutrinos, different numbers of neutrino
flavors, as well as different models of the matter density
profile. For all curves we have used $E_\nu = 1 \; {\rm GeV}$
(reasonable for atmospheric neutrinos) and $h = 0$ (completely
Earth-through-going neutrinos). The dotted line: \eq~(\ref{eq:e'em}) with
$\rho_A \simeq 7.8 \; {\rm g/cm^3}$; the dashed curve:
\eq~(\ref{eq:e2}) with $\rho_A \simeq 7.8 \; {\rm g/cm^3}$; the thin
solid curve: the Earth matter density profile with two neutrino
flavors; the thick solid curve: the Earth matter density profile with
three neutrino flavors. The two vertical dashed-dotted lines mark
the interval relevant for the Earth's atmosphere. The differences of
the slopes of the curves close to $l/L = 0$ are due to different
models of the matter density profile.}
\label{fig:2nu}
\end{figure}

In \fig~\ref{fig:2nu}, we show the relative error $\epsilon'_{e\mu}$
of the transition probability for $\nu_e \leftrightarrow \nu_\mu$ that
one makes omitting the Earth's atmosphere in the calculation of the
transition probability $P_{e\mu}$.
The dotted line is a plot of \eq~(\ref{eq:e'em}), whereas the dashed
curve is a plot of \eq~(\ref{eq:e2}). Both these plots assume a value
of the matter density parameter $A$ that corresponds to the average
value of the density of the Earth, \ie, $\rho_A \simeq 7.8 \; {\rm
g/cm^3}$. We note that \eq~(\ref{eq:e'em}) is a good approximation to
\eq~(\ref{eq:e2}) for the values of $l/L$ that are relevant for the
Earth's atmosphere. The thin and thick solid curves use the usual
mantle-core-mantle step function approximation for the Earth matter
density profile.\footnote{The mantle-core-mantle step function
approximation of the Earth matter density profile consists of three
constant matter density layers (mantle, core, and mantle) with
$\rho_{\rm mantle} = 4.5 \; {\rm g/cm^3}$ and $\rho_{\rm core} = 11.5
\; {\rm g/cm^3}$. It has been shown that the mantle-core-mantle step
function model is a very good approximation to the (realistic) Earth
matter density profile \cite{freu00}.} The thin solid curve is a plot
using two neutrino flavors, whereas the thick solid curve is a plot
using three neutrino flavors. We observe that the absolute value of
the relative error $\epsilon'_{e\mu}$ of the transition probability
$P_{e\mu}$ is of the order of 15\% - 25\% depending on the value of
the thickness of the atmosphere. This is a large error and shows that
the atmosphere cannot be neglected in trustable analyses. Increasing
the value of the vacuum mixing angle $\theta$ from $5^\circ$ to
$10^\circ$, the relative error becomes even larger $\epsilon'_{e\mu}
\sim 50\%$. In order to separate the thickness effects of the
thin density layer perturbation from the matter effects, we determine
the relative error $\epsilon'_{e\mu}$ for $A = 0$, \ie, the relative
error induced by the difference in the baseline lengths, to be
\begin{eqnarray}
&& \epsilon'_{e\mu}(L,l;0) = \frac{\sin^2 x_{21}}{\sin^2 X_{21}} + \sin 2x_{21}
\cot X_{21} \nonumber\\
&& - 2 \sin^2 x_{21} \simeq 0.0304 \approx 3\%
\label{eq:e'em_value}
\end{eqnarray}
with $L = 12742 \; {\rm km}$ and $l = 15 \; {\rm km}$. Thus, the
thickness effects are small compared with the matter effects.
Note that the above formula is independent of the vacuum mixing angle
$\theta$.

So far the present analysis for two neutrino flavors has been
conducted on the level of transition probabilities using a
monochromatic value of the neutrino energy $E_\nu$. In practice, however, the
energy resolution of an experiment must be taken into account including
energy uncertainties (or even better the analysis should be performed
on the level of neutrino event rates), since neutrinos are neither produced nor
detected with sharp energy. Another problem is attached to the
definitions of the relative errors in Eqs.~(\ref{eq:e2}) and
(\ref{eq:e'em}). In the case when $P_{e\mu}(L,l;A)$ and/or
$P_{e\mu}(L;A)$ become zero, the corresponding relative errors
$\epsilon_{e\mu}$ and $\epsilon'_{e\mu}$ go to infinity. Also when
$P_{e\mu}(L,l;A)$ and $P_{e\mu}(L;A)$ are small, we will obtain large
errors. This problem can be overcome by averaging over the neutrino
energy, which will smoothen the transition probabilities and therefore
lead to less varying errors. Thus, let us next estimate the relative 
error $\epsilon'_{e\mu}$ when the resolution of the neutrino energy
has been included. We assume for simplicity that the neutrinos
are Gaussian distributed in energy with an average neutrino energy
$\bar{E}_\nu$ and a corresponding uncertainty $\Delta E_\nu$. The Gaussian
average is defined by
\begin{eqnarray}
&& \langle \epsilon'_{e\mu} \rangle = \frac{1}{\Delta E_\nu
\sqrt{2\pi}} \nonumber\\
&& \times \int_{-\infty}^\infty \epsilon'_{e\mu}(E_\nu) e^{-\frac{(E_\nu -
\bar{E}_\nu)^2}{2 (\Delta E_\nu)^2}} \, dE_\nu.
\end{eqnarray}
The energy resolution of the Super-Kamiokande experiment is of the
order of magnitude $\Delta E_\nu \sim E_\nu$. Using this energy
resolution and an average neutrino energy $\bar{E}_\nu = 1 \; {\rm
GeV}$, we can, for example, compute the Gaussian averaged relative error
$\langle \epsilon'_{e\mu} \rangle$ for the case given in
Eq.~(\ref{eq:e'em_value}), and we obtain $\langle
\epsilon'_{e\mu}(L,l;0) \rangle \simeq 0.0181 \approx 2\%$. Indeed,
the Gaussian averaged relative error is smaller than the non-averaged one.

\begin{figure*}
\begin{center}
\includegraphics*[bb = 20 20 575 435, height=7.5cm]{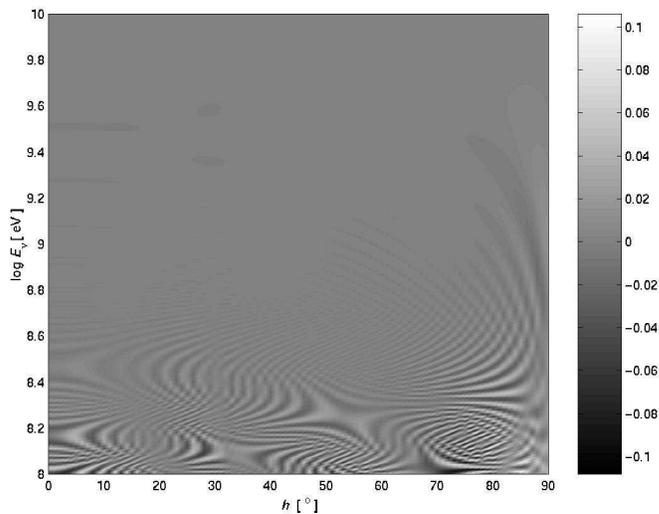}
\end{center}
\vspace{-1cm}
\caption{The difference $\Delta P_{e\mu}$ as a function of the nadir
angle $h$ and the neutrino energy $E_\nu$ for three neutrino flavors
assuming the Earth's matter density profile.}
\label{fig:3nuPem}
\end{figure*}

\begin{figure*}
\begin{center}
\includegraphics*[bb = 20 20 575 435, height=7.5cm]{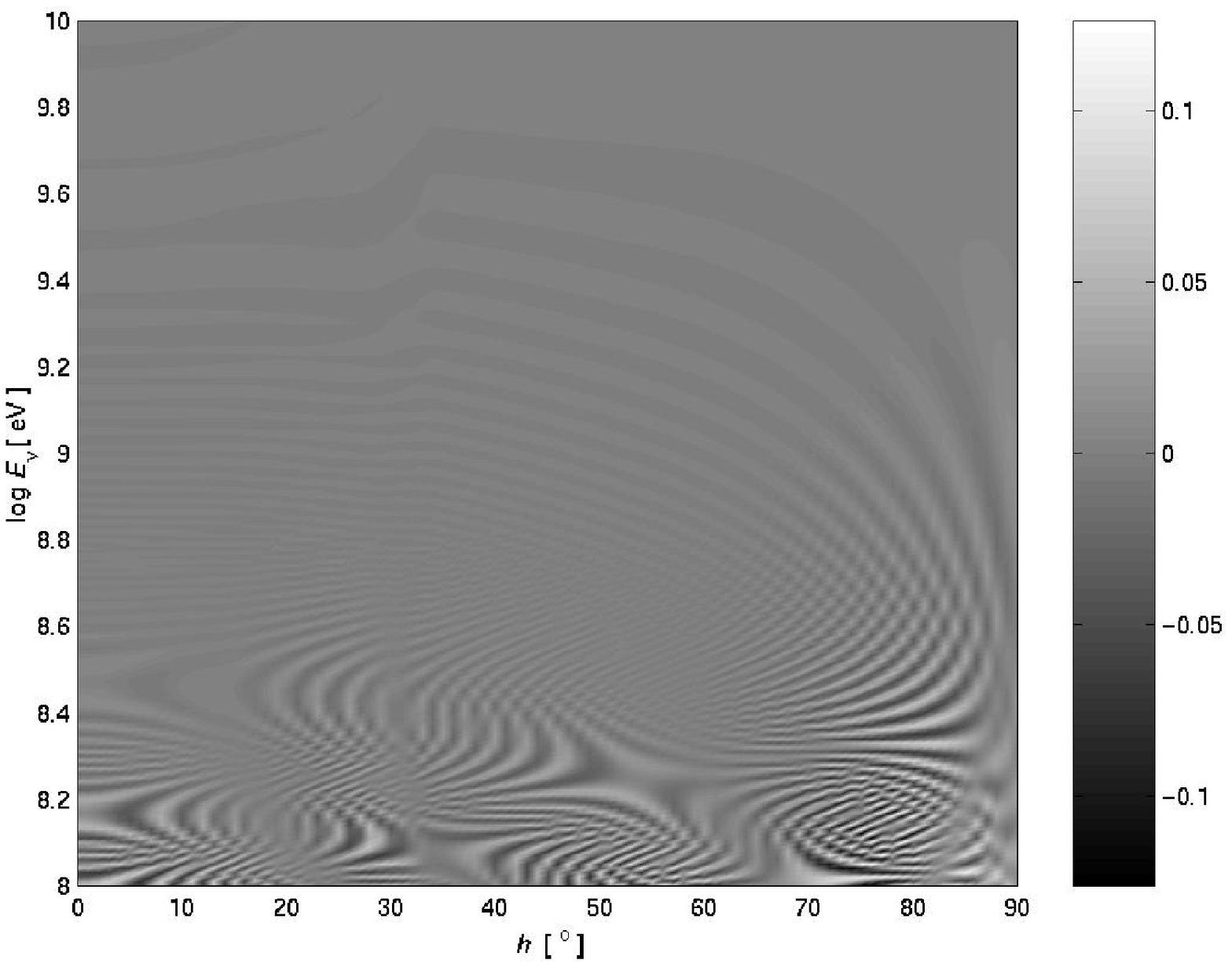}
\end{center}
\vspace{-1cm}
\caption{The difference $\Delta P_{e\tau}$ as a function of the nadir
angle $h$ and the neutrino energy $E_\nu$ for three neutrino flavors
assuming the Earth's matter density profile.}
\label{fig:3nuPet}
\end{figure*}

\begin{figure*}
\begin{center}
\includegraphics*[bb = 20 20 575 435, height=7.5cm]{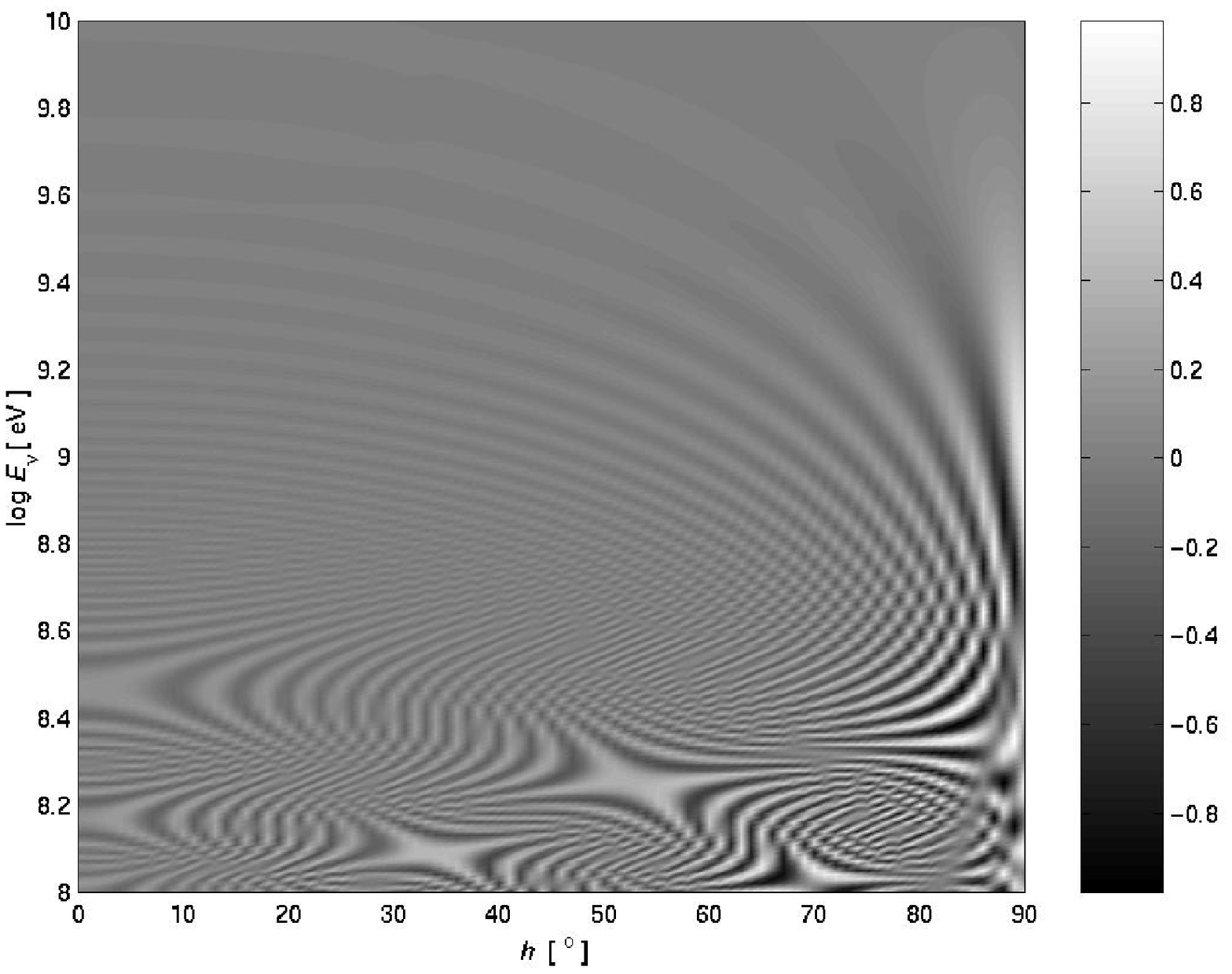}
\end{center}
\vspace{-1cm}
\caption{The difference $\Delta P_{\mu\tau}$ as a function of the nadir
angle $h$ and the neutrino energy $E_\nu$ for three neutrino flavors
assuming the Earth's matter density profile.}
\label{fig:3nuPmt}
\end{figure*}

For three neutrino flavors the most probable vacuum mixing parameters
are $\theta_{12} \simeq 45^\circ$, $\theta_{13} \simeq 5^\circ$,
$\theta_{23} \simeq 45^\circ$, $\delta_{CP} = 0$, $\Delta m_{21}^2
\simeq 3.65 \cdot 10^{-5} \; {\rm eV}^2$, and $\Delta m_{32}^2 \simeq
2.5 \cdot 10^{-3} \; {\rm eV}^2$. The values of the parameters
$\theta_{12}$ and $\Delta m_{21}^2$ correspond to the best fit point
values of the large mixing angle (LMA) solution \cite{gonz01}, whereas
the values of the parameters $\theta_{23}$ and $\Delta m_{32}^2$
correspond to the latest best fit values from the Super-Kamiokande
collaboration \cite{SuperK}. Using these values of the mixing
parameters, we have in \figs~\ref{fig:3nuPem} - \ref{fig:3nuPmt}
plotted as density plots the absolute errors of the transition
probabilities for $\nu_e \leftrightarrow \nu_\mu$, $\nu_e
\leftrightarrow \nu_\tau$, and $\nu_\mu \leftrightarrow
\nu_\tau$. These density plots also use the mantle-core-mantle step function
approximation for the Earth's matter density profile. We observe that
the errors are largest for $P_{\mu\tau}$ and smallest for
$P_{e\mu}$. Actually, for $P_{\mu\tau}$ the error can be close to
100\% depending on the nadir angle and the neutrino energy. The errors decrease
for increasing neutrino energy and naturally they increase for
increasing nadir angle. Furthermore, note that the errors are due to
both the difference in lengths of the baselines ($L' = L + l$ for
the case with the Earth's matter density profile included; $L$ for the
case without) as well as the difference in density $\Delta \rho =
\rho_A$, \ie, the errors are thickness and matter effects arising from
the thin density layer perturbation.

\section{Summary and conclusions}
\label{sec:sc}

In summary, we have derived analytically exact as well as approximate
formulas for the absolute and relative errors of the transition
probabilities for neutrino oscillations using a thin density layer
perturbation in the case of two neutrino flavors. It should be noted
that for three neutrino flavors this is not easily achieved. However,
for three neutrino flavors we have studied the influence of a thin
density layer perturbation numerically. The studies were carried out for one
of the most interesting experimental setups, \ie, atmospheric neutrinos. We
assumed the Earth's matter density profile with the Earth's atmosphere
as the thin density layer perturbation.

In conclusion, we found that the transition probabilities are rather
influenced by a thin density layer perturbation, even though its length is
at least one order of magnitude smaller than that of the matter
density profile. 
In the case of atmospheric neutrinos, using three neutrino flavors,
the absolute error $\Delta P_{\mu\tau}$ of the transition probability
$P_{\mu\tau}$ could nearly be as large as 100\% for certain realistic
nadir angles and neutrino energies. This is despite the fact that the
distance traveled by neutrinos in the Earth's atmosphere is much
shorter than the distance that they travel in the Earth's matter
density profile; at least for nadir angles $h \lesssim 84^\circ$.
Thus, it is important also to include the Earth's atmosphere in the
analyses of atmospheric neutrinos. Note that we are fully aware of
that in most such analyses this is certainly done. Furthermore, taking
energy resolution into account will definitely reduce the
errors of the transition probabilities. The purpose of this 
work was to investigate and determine the actual effects of the
Earth's atmosphere being in principle just a thin density layer
perturbation in comparison with the Earth's matter density profile itself.  

Finally, note that this paper is in principle the ``counterpart'' to the
study performed in \Ref~\cite{akhm01}, in which the distance traveled
by neutrinos in matter is assumed to be short compared with the
distance in vacuum.

\section*{Acknowledgements}

I would like to thank Evgeny Akhmedov, Martin Freund, Walter Winter,
and the referee for useful discussions and comments and Martin Freund,
H{\aa}kan Snellman, and Walter Winter for valuable advices.

This work was supported by the Swedish Foundation for International
Cooperation in Research and Higher Education (STINT), the Wenner-Gren
Foundations, and the ``Sonderforschungsbereich 375
f{\"u}r Astro-Teilchenphysik der Deutschen Forschungsgemeinschaft''.

\bibliographystyle{h-elsevier}
\bibliography{references}

\end{document}